\documentclass[referee]{raa}            

\usepackage{graphicx,times}             

\begin{document}

   \title{{\em Fermi}/LAT observations of Lobe-dominant Radio Galaxy 3C 207 and Possible Radiation Region of the Gamma-Rays }

   \volnopage{Vol.0 (200x) No.0, 000--000}      
   \setcounter{page}{1}          

   \author{Sheng-Chu Guo
      \inst{1}
   \and Hai-Ming Zhang
      \inst{1}
   \and Jin Zhang
      \inst{2\dag}
   \and En-Wei Liang
      \inst{1}
   }

   \institute{Guangxi Key Laboratory for Relativistic Astrophysics, Department of Physics, Guangxi University, Nanning 530004, China\\
        \and
          Key Laboratory of Space Astronomy and Technology, National Astronomical Observatories, Chinese Academy of Sciences, Beijing 100012, China; jinzhang@bao.ac.cn   \\
   }

   \date{Received~~201X month day; accepted~~201X~~month day}

\abstract{3C 207 is a lobe-dominant radio galaxy with one sided jet and the bright knots in kpc-Mpc scale were resolved in the radio, optical and X-ray bands. It was confirmed as a $\gamma$-ray emitter with {\em Fermi}/LAT, but it is uncertain whether the $\gamma$-ray emission region is the core or knots due to the low spatial resolution of {\em Fermi}/LAT. We present an analysis of its \emph{Fermi}/LAT data in the past 9 years. Different from the radio and optical emission from the core, it is found that the $\gamma$-ray emission is steady without detection of flux variation over 2$\sigma$ confidence level. This likely implies that the $\gamma$-ray emission is from its knots. We collect the radio, optical, and X-ray data of knot-A, the closest knot from the core at 1.$\arcsec$4, and compile its spectral energy distribution (SED). Although the single-zone synchrotron+SSC+IC/CMB model by assuming knot-A at rest can reproduce the SED in the radio-optical-X-ray band, the predicted $\gamma$-ray flux is lower than the LAT observations and the derived magnetic field strength deviates the equipartition condition with 3 orders of magnitude. Assuming that knot-A is relativistically moving, its SED from radio to $\gamma$-ray bands would be well represented with the single-zone synchrotron+SSC+IC/CMB model under the equipartition condition. These results likely suggest that the $\gamma$-ray emission may be from knot-A via the IC/CMB process and the knot should have relativistical motion. The jet power derived from our model parameters is also roughly consistent with the kinetic power estimated with the radio data.
}

   \authorrunning{Sheng-Chu Guo et al. }            
   \titlerunning{The $\gamma$-ray Emission of 3C 207 }  

   \maketitle
%
%
\section{Introduction}           
\label{sect:intro}

Jet substructures, including knots and lobes with hotspots, in kpc-Mpc-scale are resolved in the radio, optical, and X-ray bands for some powerful jets of active galactic nuclei (AGNs), especially for radio galaxies (Harris \& Krawczynski 2006 for a review). It is interesting that the radio lobes of Cen A and Fornax A  were also detected in the $\gamma$-ray band with \emph{Fermi}/LAT, likely suggesting that these substructures are powerful particle accelerators (Abdo et al. 2010a; Sun et al. 2016; McKinley et al. 2015; Ackermann et al. 2016). The radio emission of these substructures, as well as the optical emission of the substructures in some sources, is believed to be produced by the synchrotron radiation of relativistic electrons (Tavecchio et al. 2000; Hardcastle et al. 2004; Harris \& Krawczynski 2006; Sambruna et al. 2006; Zhang et al. 2009, 2010). For the X-ray emission of most substructures, the spectrum is harder than the radio-optical spectrum, showing that it is not an extrapolation from radio-optical band. A new component needs to be involved for producing the X-ray emission and the emission of higher energy band, such as the synchrotron emission of a second population of very-high-energy electrons (Zhang et al. 2009; Meyer et al. 2015; Zargaryan et al. 2017; Sun et al. 2017), the synchrotron self-Compton process (SSC, Hardcastle et al. 2004; Kataoka \& Stawarz 2005; Zhang et al. 2010), and the inverse Compton scattering of cosmic microwave background photons (IC/CMB, Tavecchio et al. 2000; Sambruna et al. 2006; Zhang et al. 2010; Zhang \& Li 2010). Since these models predict different $\gamma$-ray emission features (e.g., Zhang et al. 2010), the observations with \emph{Fermi}/LAT then may play strong constraints on these radiation models (e.g., Zhang et al. 2010; Meyer et al. 2015).

3C 207 is an FR II radio galaxy at redshift of $z=0.684$  with lobe-dominant radio morphology (Hough \& Readhead 1989). The ratio of core to extended lobe emission at 8.4 GHz is 0.82 (Hough et al. 2002). Its pc-scale jet structure was resolved by the Very Long Baseline Array (VLBA) at 8.4 GHz and is distinctly curved with a bright inner knot misaligned from the jet axis defined by the outer weaker jet features (Hough et al. 2002). It has a powerful one sided radio and X-ray large-scale jet and the extended X-ray emission coincident with the western radio lobe is also detected in the opposite direction (Brunetti et al. 2002). Three radio knots at 1.$\arcsec$4 ( 9.94 kpc, knot-A), 4.$\arcsec$6 (32.66 kpc, knot-B), and 6.$\arcsec$5 (46.15 kpc, knot-C, maybe a hotspot) are resolved and detected in the \emph{Chandra} image. Among the three knots only knot-A was detected in the optical band with the \emph{Hubble Space Telescope} (HST, Sambruna et al. 2004). 3C 207 was confirmed as a $\gamma$-ray emitter by the \emph{Fermi}/LAT observations (Abdo et al. 2010b), but the emitting region is uncertain whether it is the core or the extended substructures since LAT is inadequate to spatially resolve the location of $\gamma$-ray emission.

This paper presents the analysis of the {\em Fermi}/LAT observations for 3C 207 in the past 9 years in order to investigate the radiation region and the radiation mechanisms of the gamma-rays. The observations and data reduction are reported in Section 2. Data analysis and the broadband SED modeling of knot-A are given in Section 3. Discussion and Conclusions are described in Section 4. Throughout, $H_0=71$ km s$^{-1}$ Mpc$^{-1}$, $\Omega_{m}=0.27$, and $\Omega_{\Lambda}=0.73$ are adopted.

\section{\emph{Fermi}/LAT Observations and Data Reduction}

3C 207 was detected in the $\gamma$-ray band since the launch of {\em Fermi} mission. We download the LAT data of 3C 207 covering from 2008 August 4 (Modified Julian Day, MJD 54682) to 2017 July 20 (MJD 57954) from the Fermi data archive (Pass 8 data). The temporal coverage of the LAT observations is $\sim$9 years. The well-sampling of the LAT data provides a good opportunity to investigate both the short- and long-term variability of sources. The standard analysis tool \emph{gtlike/pyLikelihood}, which is part of the \emph{Fermi} Science Tool software package (ver. v10r0p5), and the P8R2-SOURCE-V6 set of instrument response functions (IRFs) were used to analyze the LAT data. The significance of the $\gamma$-ray signal from the source is evaluated with the maximum-likelihood test statistic. The events with energies from 0.1 to 100 GeV are selected from the region of interest (ROI) with radius of 10$^{\circ}$, centered at the position of 3C 207 (R.A.=130.198, decl.=13.2065). All point sources in the third \emph{Fermi}/LAT source catalog (Acero et al. 2015) located in the ROI and an additional surrounding 10$^{\circ}$ wide annulus (called ``source region") were modeled in the fits. In the model file, the spectral parameters for sources lying within the ROI were kept free and for sources lying within ``source region" were fixed. The isotropic background, including the sum of residual instrumental background and extragalactic diffuse $\gamma$-ray background, was fitted with a model derived from the isotropic background at high Galactic latitude, i.e., ``iso-P8R2-SOURCE-V6-v06.txt", and the Galactic diffuse GeV emission was modeled with ``gll-iem-v06.fits". In order to eliminate the contamination from the $\gamma$-ray-bright Earth limb, the events with zenith angle larger than 95$^{\circ}$ are excluded. The spectral analysis in the energy range of 0.1--100 GeV was performed by using the \textit{unbinned likelihood analysis}. A power law (PL) model , i.e., $dN(E)/dE = N_0E^{-\Gamma_{\gamma}}$, is used to fit the spectrum accumulated in each time bin.

\section{Data analysis and SED Modeling}

The derived light curve in time-bins of 30 days is shown in Figure \ref{LC}. No significant flux variation in the $\gamma$-ray band is observed. In the long-term variation point of view, the $\gamma$-ray flux is steady with small flux variation around $3.15\times 10^{-8}$ photons cm$^{-2}$ s$^{-1}$, where $\bar{F}=3.15\times 10^{-8}$ photons cm$^{-2}$ s$^{-1}$ is the mean of these data points in Figure \ref{LC} and indicates the average flux in the past 9 years. In the time-scale of months, we measure the confidence level of flux variation with $(F_{\max}-\bar{F})/\sigma_{F_{\max}}$, where $F_{\max}$ and $\sigma_{F_{\max}}$ are the highest flux and its error in the past 9 years. The confidence lever is 1.4, being smaller than 2$\sigma$\footnote{Liao et al. (2015) reported that 3C 207 has significant variability around MJD 54750 on a timescale of months with a six-year light curve in time-bins of 7 months. Besides the different `Pass data' and time-bin, their method to calculate the long-term average flux is also different from our work. Note that there is another high flux point after MJD 57000 in Figure \ref{LC}. Hence, we obtain a higher 9-year average flux than their 6-year average flux.}.

\begin{figure*}
\includegraphics[angle=0,scale=0.43]{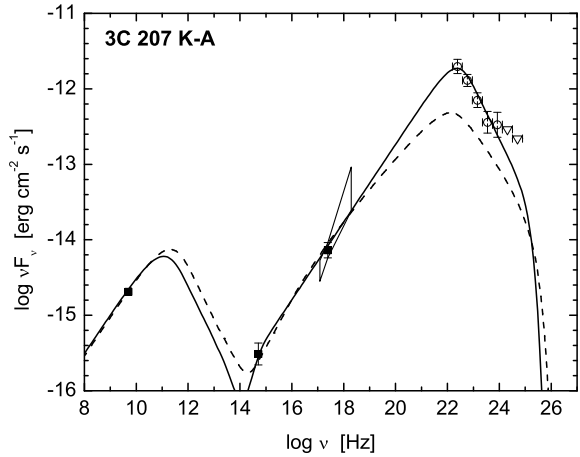}
\caption{Observed SED and model fitting lines with the single-zone synchrotron+SSC+IC/CMB models for knot-A. The dashed line and solid line are for considering knot-A at rest and the beaming effect under equipartition condition, respectively. The fitting parameters for considering knot-A at rest are $R=1.1\times10^{22}$ cm, $B=(1.4\pm0.4)~\mu$G, $N_0=(0.01\pm0.002)$ cm$^{-3}$, $p_1=1.97$, $p_2=4.4$, $\gamma_{\rm min}=200$, $\gamma_{\rm b}=(2.5\pm1.6)\times10^5$, and $\gamma_{\rm max}=5\times10^6$. The fitting parameters for considering the beaming effect under equipartition condition are $R=1.1\times10^{22}$ cm, $\Gamma=\delta=13.3\pm1.5$, $B=5.9\pm1.7~\mu$G, $N_0=(1.4\pm0.8)\times10^{-7}$ cm$^{-3}$, $p_1=1.97$, $p_2=4.4$, $\gamma_{\rm min}=3$, $\gamma_{\rm b}=(2.5\pm0.6)\times10^4$, and $\gamma_{\rm max}=5\times10^5$. The opened symbols are the average spectrum of the \emph{Fermi}/LAT observations, where the down-triangles in the $\gamma$-ray band indicate the upper limits. }\label{sed}
\end{figure*}

The steady $\gamma$-ray emission may imply that the emission is from the large-scale substructures of jet. We therefore further explore whether the $\gamma$-rays are from the knots. Three knots, knot-A, knot-B, and knot-C, were resolved in the radio and X-ray bands for 3C 207, but only knot-A, the closest knot from the core at 1.$\arcsec$4, was also detected in the optical band (Sambruna et al. 2004). We compile the broadband SED in the radio, optical, and X-ray bands of knot-A, as well as the average spectrum in the $\gamma$-ray band, which is shown in Figure \ref{sed}. The X-ray spectral index is $0.1\pm0.3$, which is harder than the radio-to-optical spectral index, i.e., $1.16\pm0.02$ (Sambruna et al. 2004). The radio emission of knots is believed to be produced by the synchrotron radiations of the relativistic electrons in knots. The X-ray emission with a harder spectrum than that in the radio-optical band cannot be produced by the same electron population via the synchrotron radiations. Therefore, the IC processes would be involved to explain the X-ray emission and the higher energy emission.

We model the SED of knot-A with the single-zone synchrotron+IC radiation models. The emitting region is assumed to be a sphere with radius of $R$, magnetic field strength of $B$, and a bulk Lorentz factor $\Gamma$ at an angle $\theta$ with respect to the line of sight. Since the optical flux of knot-A is extracted within a circle of radius 0.$\arcsec$5 (Sambruna et al. 2004), we fix the dimension of $R=1.1\times10^{22}$ cm corresponding to this angular size. The radiation region is homogeneously filled by the relativistic electrons, and the electron distribution is assumed as a broken power law, i.e.,
\begin{equation}
N(\gamma )= N_{0}\left\{ \begin{array}{ll}
\gamma ^{-p_1}  &  \mbox{ $\gamma_{\rm min}\leq\gamma \leq \gamma _{\rm b}$}, \\
\gamma _{\rm b}^{p_2-p_1} \gamma ^{-p_2}  &  \mbox{ $\gamma _{\rm b} <\gamma <\gamma _{\rm max} $.}
\end{array}
\right.
\end{equation}
$p_1$ and $p_2$ can be constrained with the observed spectral indices $\alpha_{\rm ox}$ and $\alpha_{\gamma}$, where $\alpha_{\rm ox}$ is broadband spectral index in the optical and X-ray bands and $\alpha_{\gamma}$ is spectral index in $\gamma$-ray band. The Klein-Nashina (KN) effect and the absorption of high energy $\gamma$-ray photons by extragalactic background light (Franceschini et al. 2008) are also considered in our calculations.

Two photon fields are involved for the IC processes in the model calculations, i.e., synchrotron radiation photon field and the CMB photon field. The energy density of the synchrotron radiation photon field is estimated as $U_{\rm syn}=\frac{L_{\rm s}}{4\pi R^2c}\simeq\frac{5L_{\rm 5GHz}}{4\pi R^2c}=4.6\times10^{-13}$ erg cm$^{-3}$, where we assume that the synchrotron luminosity is five times of $L_{\rm 5GHz}$ and $L_{\rm 5GHz}$ is the luminosity of knot-A at 5 GHz. The CMB energy density in the comoving frame is derived as $U^{'}_{\rm CMB}=\frac{4}{3}\Gamma^2U_{\rm CMB}(1+z)^4$ , where $U_{\rm CMB}=4.2\times10^{-13}$ erg cm$^{-3}$ (Dermer \& Schlickeiser 1994). It is found that $U_{\rm syn}$ and $U_{\rm CMB}$ are comparable. Therefore, two photon fields are taken into account for modeling the SED of knot-A.

As illustrated in Figure \ref{sed}, the X-ray spectral index places a strong constraint on the SED modeling, and the predicted $\gamma$-ray flux by the models should not exceed the observational flux of the \emph{Fermi}/LAT. We first assume that knot-A is at rest to model the SED of knot-A. The result is shown in Figure \ref{sed}. It is found that the SED from radio to X-ray bands is well reproduced by the model, but the predicted $\gamma$-ray flux by the model is lower than the \emph{Fermi}/LAT observations. During the SED modeling, $\gamma_{\rm min}$ and $\gamma_{\rm max}$ cannot be constrained, and thus they are fixed, i.e., $\gamma_{\rm min}=200$ as reported in Zhang et al. (2010) and $\gamma_{\rm max}=5\times 10^6$. The derived other parameters are $B=(1.4\pm0.4)~\mu$G, $N_0=(0.01\pm0.002)$ cm$^{-3}$, $\gamma_{\rm b}=(2.5\pm1.6)\times10^5$, $p_1=1.97$, and $p_2=4.4$. The corresponding equipartition magnetic field strength is $B_{\rm eq}=0.0014$ G, which is much higher than the derived magnetic field strength in this scenario. This issue of the SSC model has been widely reported (Harris \& Krawczynski 2006  for a review).

We then model the SED of knot-A under the equipartition condition by considering the beaming effect with an assumption of $\Gamma=\delta$. In this scenario, we try to constrain other fitting parameters by taking the $\gamma$-ray data into account in SED modeling. A small value of $\gamma_{\rm min}$ (smaller than 3) is needed to model the optical data, hence we take $\gamma_{\rm min}=3$. The $\gamma_{\rm max}$ is still fixed at a large value, i.e., $\gamma_{\rm max}=5\times 10^5$, which does not significantly affect our results. As illustrated in Figure \ref{sed}, the SED from radio to $\gamma$-ray bands can be well represented by the model. The derived parameters are $\delta=13.3\pm1.5$, $B_{\rm eq}=(5.9\pm1.7)~\mu$G, $N_0=(1.4\pm0.8)\times 10^{-7}$ cm$^{-3}$, and $\gamma_{\rm b}=(2.5\pm0.6)\times 10^4$.

\section{Conclusions and Discussion}
We have presented an analysis of the \emph{Fermi}/LAT data in the past 9 years for 3C 207. It is found that the $\gamma$-ray emission is steady without detection of flux variation over 2$\sigma$ confidence level. We suspect that the $\gamma$-ray emission is from its knots in the kpc-scale. By collecting data from the literature we compile the SED in the radio, optical, and X-ray data of knot-A in 3C 207. Although the single-zone synchrotron+SSC+IC/CMB model by assuming that knot-A is at rest can reproduce its SED in the radio-optical-X-ray band, the predicted $\gamma$-ray flux is lower than the LAT observations and the derived magnetic field strength deviates the equipartition condition with 3 orders of magnitude. Assuming that the knot is relativistically moving, the SED of knot-A in the radio to X-ray bands as well as the $\gamma$-ray emission would be well represented under the equipartition condition. These results likely suggest that the $\gamma$-ray emission may be from knot-A via IC/CMB process and knot-A should have relativistical motion.

\begin{figure*}
\includegraphics[angle=0,scale=0.4]{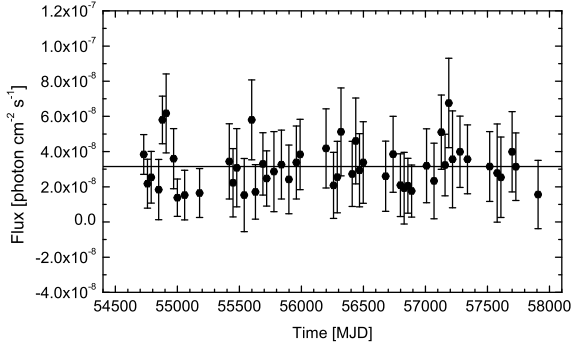}

\caption{Global lightcurve of 3C 207 in the GeV band in time-bins of 30 days, from MJD 54682 to MJD 57954. The solid horizontal line indicates the average flux of the \emph{Fermi}/LAT observations.}\label{LC}
\end{figure*}

Significant flux variations in the radio and optical bands in time-scale of days from the core region of 3C 207 were detected\footnote{http://crts.caltech.edu/ and http://sma1.sma.hawaii.edu/smaoc.html}. As shown in Figure \ref{LC}, we do not find any flux variation in such a comparable timescale in a confidence level of 2$\sigma$. In addition, the $\gamma$-ray flux from the core region is usually correlated with the radio and and optical fluxes from the same region (e.g., Jorstad et al. 2013; Zhang et al. 2017). The non-correlation between the $\gamma$-rays with the radio and optical emission for 3C 207 may indicate that the $\gamma$-rays are not from the core region. These factors further support that the $\gamma$-ray emission of 3C 207 may be dominated by its knot-A. In addition, the steady $\gamma$-ray emission of 3C 207 may be contributed by both the core and knots. And thus the fluctuations of the core emission would be concealed if the flux levels of emission from the core and knots are comparable.

An apparent superluminal speed of $\beta_{\rm app}\sim$ 10$c$ in pc-scale jet for 3C 207 has been reported by Hough et al. (2002) and Lister et al. (2016). Note that knot-A is the closest knot from the core region (Sambruna et al. 2004). We obtain $\Gamma\sim 13$ by modeling SED for knot-A, suggesting that the knot still relativistically moves. The one-sided jets in AGNs clearly imply that the kpc-scale jets are at least mildly relativistic (Bridle et al. 1994), hence our results are consistent with the observed one-sided jet in 3C 207, and the $\gamma$-ray emission is produced by the IC/CMB process.

One issue for the IC/CMB model to produce the high energy emission in the large-scale jet substructures is that it may result in the so-called ``super-Eddington" jet powers as suggested by some authors (e.g., Dermer \& Atoyan 2004; Uchiyama et al. 2006; Meyer et al. 2015). We calculate the jet power of knot-A on the basis of the model parameters derived from our SED fit in case of knot-A being relativistic motion. Since the contribution of protons to jet power is not exactly known (see also Zargaryan et al. 2017), we assume that the jet consists of electrons and magnetic fields only, i.e., $P_{\rm jet}=\pi R^2 \Gamma^2c(U_{\rm e}+U_B)$. We obtain $P_{\rm jet}=5.77\times10^{45}$ erg s$^{-1}$. We also estimate the kinetic power of knot-A with the radio data, i.e., $L_{\rm k}=3\times10^{45}f^{3/2}L_{151}^{6/7} \quad\rm erg\ s^{-1}$ (Willott et al. 1999), where $L_{151}$ is the 151 MHz radio flux in units of $10^{28}$ W Hz$^{-1}$ sr$^{-1}$ and taken from Arshakian et al. (2010, see the NASA/IPAC Extragalactic Database). We take $f=10$ (Cattaneo \& Best 2009) and obtain $L_{\rm k}=2.49\times10^{46}$ erg s$^{-1}$. It is found that $P_{\rm jet}$ derived from our model parameters is lower than $L_{\rm k}$\footnote{Willott et al. (1999) reported that the bulk kinetic power of the jet can be estimated with the extended radio luminosity at 151 MHz under the assumption that the minimum stored energy is required in the extended regions to produce the observed synchrotron luminosity. If the beaming effect exist in large-scale jets, the derived kinetic power with the extended radio luminosity at 151 MHz would be overestimated.}.

\begin{acknowledgements}
We thank the anonymous referee for his/her valuable suggestions. This work is supported by the National Natural Science Foundation of China (grants 11573034, 11533003 and U1731239), the National Basic Research Program (973 Programme) of China (grant 2014CB845800), and En-Wei Liang acknowledges support from the special funding from the Guangxi Science Foundation for Guangxi Distinguished Professors (Bagui Yingcai \& Bagui Xuezh; 2017AD22006).

\end{acknowledgements}


\end{document}